\documentstyle[festkoer,psfig]{vieweg}

\author{Ralf Bulla}

\Kurzautor{Bulla}

\title{The Numerical Renormalization Group Method for correlated
electrons}

\Kurztitel{The Numerical Renormalization Group Method for correlated
electrons}

\Adresse{Theoretische Physik III, Elektronische Korrelationen und Magnetismus,
Universit\"at Augsburg, 86135 Augsburg }

\begin{document}

\Titel

\begin{abstract}
The Numerical Renormalization Group method (NRG) has been developed
by Wilson in the 1970's to investigate the Kondo problem. The
NRG allows the  non-perturbative calculation of static and dynamic
properties for a variety of impurity models. In addition,
this method has been recently generalized to lattice models within the
Dynamical Mean Field Theory. This paper gives a brief historical overview 
of the development of the NRG and discusses its application to the
Hubbard model; in particular the results for the Mott metal-insulator
transition at low temperatures.

\end{abstract}

\section{The Numerical Renormalization Group and the Kondo problem}

The application of renormalization group (RG) ideas in the physics of condensed
matter has been strongly influenced by the work of Wilson \cite{Wil75}.
His `theory for critical phenomena in connection
with phase transitions'  has been awarded the Nobel prize in
physics in 1982 \cite{Nobel}. This paper deals with 
one aspect in the work of Wilson: the numerical 
renormalization group (NRG) method for the investigation of the Kondo problem.

The history of the Kondo problem \cite{hew1} goes back to the 1930's when a 
resistance minimum was found at very low temperatures in seemingly
pure metals \cite{deHaas}. This minimum, and the strong increase of the resistance 
$\rho(T)$
on further lowering the temperature, has been later found to be caused
by magnetic impurities (such as iron). Kondo successfully explained the
resistance minimum within a perturbative calculation for the $s$-$d$-
(or Kondo-) model \cite{Kondo}, a model for magnetic impurities in metals.
However, Kondo's result implies a divergence of $\rho(T)$ for
$T\to 0$, in contrast to the saturation found experimentally.
It became clear that this shortcoming  is due
to the perturbative approach used by Kondo.

An important step towards a solution of this problem (the `Kondo problem')
has been the scaling approach by Anderson \cite{Anderson}. 
By successively eliminating
high energy states, Anderson showed that the coupling $J$ in the effective
low energy model diverges. However, the derivation only holds 
within perturbation theory
in $J$ and is therefore not necessarily valid in the large $J$ limit.
A diverging coupling between impurity
and conduction electrons corresponds to a perfect screening of the
impurity spin; the magnetic moment therefore vanishes for $T\to 0$
and the resistivity no longer diverges.
This result has been finally
verified by Wilson's NRG, as will be discussed below. 

In the following,
some details of the NRG method are explained in the context of the
single impurity Anderson model \cite{SIAM}  (Wilson 
originally set up the RG transformation for the Kondo model, but
the details of the NRG are essentially the same for both models 
\cite{Wil75,Kri80}). 
The Hamiltonian of this
model is given by
\begin{eqnarray}
  H &=&   \sum_{\sigma} \varepsilon_{\rm f} f^\dagger_{\sigma}
                             f_{\sigma}
                 + U  f^\dagger_{\uparrow} f_{\uparrow}
                       f^\dagger_{\downarrow} f_{\downarrow}
                \nonumber \\
           & & + \sum_{k \sigma} \varepsilon_k c^\dagger_{k\sigma}
c_{k\sigma}
            +  \sum_{k \sigma} V
           \Big( f^\dagger_{\sigma} c_{k \sigma}
               +   c^\dagger_{k\sigma} f_{\sigma} \Big). 
    \label{eq:siam}
\end{eqnarray}
In the model (\ref{eq:siam}), $c_{k\sigma}^{(\dagger)}$ denote 
annihilation
(creation) operators for band states with 
spin $\sigma$ and energy $\varepsilon_k$,
$f_{\sigma}^{(\dagger)}$
those for impurity states with spin $\sigma$ and energy 
$\varepsilon_{\rm f}$. The
Coulomb interaction for two electrons at the impurity site is given by
$U$ and
both subsystems are coupled via a hybridization $V$.

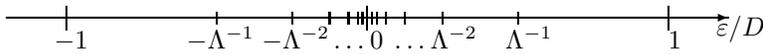
\begin{figure}[ht]

\unitlength0.8cm
\begin{picture}(14,2.2)

\put(1,1){\vector(1,0){12}}
\put(2,0.8){\line(0,1){0.4}}
\put(12,0.8){\line(0,1){0.4}}
\put(7,0.8){\line(0,1){0.4}}
\put(1.8,0.5){$-1$}
\put(12,0.5){$1$}
\put(7.05,0.5){$0$}
\put(12.8,0.7){$\varepsilon/D$}

\put(4.5,0.9){\line(0,1){0.2}}
\put(5.75,0.9){\line(0,1){0.2}}
\put(6.375,0.9){\line(0,1){0.2}}
\put(6.6875,0.9){\line(0,1){0.2}}
\put(6.8438,0.9){\line(0,1){0.2}}
\put(6.9219,0.9){\line(0,1){0.2}}

\put(4.,0.5){$-\Lambda^{-1}$}
\put(5.25,0.5){$-\Lambda^{-2}\dots$}
\put(7.45,0.5){$\dots$}
\put(8.05,0.5){$\Lambda^{-2}$}
\put(9.3,0.5){$\Lambda^{-1}$}

\put(7.0781,0.9){\line(0,1){0.2}}
\put(7.1562,0.9){\line(0,1){0.2}}
\put(7.3125,0.9){\line(0,1){0.2}}
\put(7.625,0.9){\line(0,1){0.2}}
\put(8.25,0.9){\line(0,1){0.2}}
\put(9.5,0.9){\line(0,1){0.2}}

\end{picture}
\label{fig:disc}
\caption[fig1]{Logarithmic discretization of the conduction band}
\end{figure}

The first step to set up the RG-transformation is a logarithmic
discretization of the conduction band (see Fig. \ref{fig:disc}): 
the continuous
conduction band is divided into (infinitely many) intervals
$[\xi_{n+1},\xi_n]$ and $[-\xi_{n},-\xi_{n+1}]$ with 
$\xi_n = D \Lambda^{-n}$ and $n=0,1,2,\ldots$.
$D$ is the half-bandwidth of the conduction band and $\Lambda$ the
NRG-discretization parameter (typical values used in the calculations
are $\Lambda = 1.5,\ldots,2$). The conduction band states in each 
interval are then replaced by a {\em single} state. Although this 
approximation by a discrete set of states involves some coarse graining
at higher energies, it captures arbitralily small energies near 
the Fermi level.

In a second step, the discrete model is mapped on a semi-infinite chain
form described by the hamiltonian (see also Fig. 2):

\newpage
\begin{eqnarray}
  H &=& \sum_{\sigma} \varepsilon_{\rm f} f^\dagger_{-1 \sigma}
                             f_{-1 \sigma}
                 + U  f^\dagger_{-1 \uparrow} f_{-1 \uparrow}
                       f^\dagger_{-1 \downarrow} f_{-1 \downarrow}
                \nonumber \\
           &+& \sum_{\sigma n=-1}^\infty \varepsilon_n \Big(
                f^\dagger_{n \sigma} f_{n+1 \sigma}
              +  f^\dagger_{n+1 \sigma} f_{n \sigma}  \Big)
\label{eq:chain}
\end{eqnarray}
Here, the impurity operators are written as $f^{(\dagger)}_{-1 \sigma}$
and the conduction band states as $f^{(\dagger)}_{n \sigma}$ with 
$n=0,1,2,\dots$. Due to the logarithmic discretization, the hopping
matrix elements decrease as $\varepsilon_n\propto\Lambda^{-n/2}$.
This can be easily understood by considering a discretized  
conduction band with a finite number of states $M$ (with $M$ even).
The lowest energy scale is, according to Fig. \ref{fig:disc} given by
$\approx D\Lambda^{-M/2}$. This discrete model is mapped onto
a semi infinite chain with the same number of conduction electron degrees
of freedom, $M$. The only way to generate the low energy scale
$\approx D\Lambda^{-M/2}$ is now due to the hopping matrix elements
$\varepsilon_n$ so that they have to fall of with the square root
of Lambda.

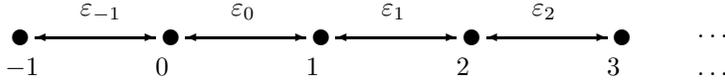
\begin{figure}[ht]
\begin{center}
\unitlength1cm
\begin{picture}(14,2)

\put(1,1){\circle*{0.2}}
\put(3,1){\circle*{0.2}}
\put(5,1){\circle*{0.2}}
\put(7,1){\circle*{0.2}}
\put(9,1){\circle*{0.2}}

\put(2,1){\vector(1,0){0.8}}
\put(2,1){\vector(-1,0){0.8}}
\put(4,1){\vector(1,0){0.8}}
\put(4,1){\vector(-1,0){0.8}}
\put(6,1){\vector(1,0){0.8}}
\put(6,1){\vector(-1,0){0.8}}
\put(8,1){\vector(1,0){0.8}}
\put(8,1){\vector(-1,0){0.8}}

\put(0.8,0.5){$-1$}
\put(1.8,1.3){$\varepsilon_{-1}$}
\put(2.8,0.5){$0$}
\put(3.8,1.3){$\varepsilon_{0}$}
\put(4.8,0.5){$1$}
\put(5.8,1.3){$\varepsilon_{1}$}
\put(6.8,0.5){$2$}
\put(7.8,1.3){$\varepsilon_{2}$}
\put(8.8,0.5){$3$}

\put(10,0.5){$\dots$}
\put(10,1.0){$\dots$}

\end{picture}
\end{center}
\label{chain}
\caption[fig1]{Semi-infinite chain form of the single impurity Anderson
model}
\end{figure}

This means that, in going along the chain, the system evolves
from high energies (given by $D$ and $U$) to arbitralily
low energies (given by $D\Lambda^{-M/2}$). The renormalization group
transformation is now set up in the following way.

We start with the solution of the isolated impurity, that is
the knowledge of all eigenstates, eigenenergies and matrix elements.
The first step of the renormalization group
transformation is to add the first conduction electron site, set up
the hamiltonian matrices for the enhanced Hilbert space, and obtain
the information for the new eigenstates, eigenenergies and matrix elements
by diagonalizing these matrices. This procedure is then iterated.
An obvious problem occurs after only a few steps of 
the iteration. The Hilbert space grows as $4^N$, which makes it impossible
to keep all the states in the calculation. Wilson therefore devised
a very simple truncation procedure in which only those states (typically
a few hundred) with the lowest energies are kept. This truncation scheme
is very successful but relies on the fact that the hopping matrix
elements are falling of exponentially. High energy states therefore
do not change the very low frequency behaviour and can be neglected.

\begin{figure}
\hspace*{2.3cm}\psfig{figure=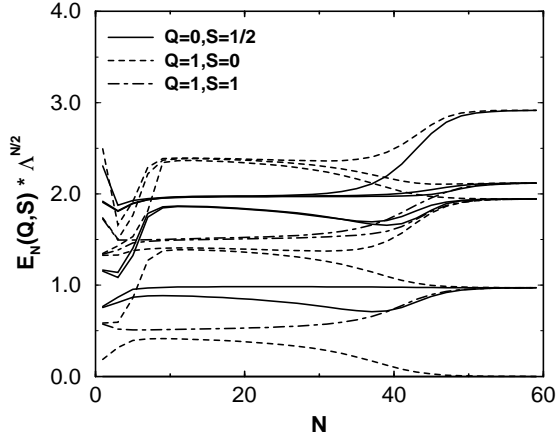,width=8cm}
\label{fig:flow}
\caption[]{Flow diagram for the lowest lying energy levels for the single
impurity Anderson model with $\varepsilon_f = -0.2$, $U=0.4$ and $\Delta=0.015$.}
\end{figure}

This procedure gives for each cluster a set of eigenenergies and matrix
elements from which a number of physical properties can be derived
(this will be illustrated for the calculation of the spectral function in the
next section). The eigenenergies itself show the essential physics
of the Kondo problem: Fig. 3 shows the dependence of the lowest
lying energy levels on the length of the chain (the energies are scaled
by a factor $\Lambda^{N/2}$). The system is first approaching an unstable
fixed point at $N\approx 10 - 20$ (the Local Moment fixed point) and is then flowing
to a stable fixed point for $N>50$ (the Strong Coupling fixed point).
By analyzing the structure of the Strong Coupling fixed point and by 
calculating perturbative corrections about it, Wilson 
(for the Kondo model \cite{Wil75}) and 
Krishnamurthy, Wilkins and Wilson (for the single
impurity Anderson model \cite{Kri80}) found that
\begin{itemize}
  \item[i)] right at the fixed point, the impurity spin is completely screened;
  \item[ii)] on approaching the fixed point, the thermodynamic properties
           are Fermi-liquid like; i.e. the magnetic susceptibility $\chi(T)$
           approaches a constant value for $T\to 0$ and the specific heat
           $C = \gamma T$ is linear in $T$ for $T\to 0$; the ratio 
           $R=\chi/\gamma$ is known as the Wilson ratio and takes the
           universal value $R=2$ in the Kondo model;
\end{itemize}

\section{Developments and Applications of the NRG method}

The NRG approach decribed so far has two main advantages:
it is non-per\-tur\-ba\-tive and can deal with arbitrary values of $U$
(simply because the impurity part is diagonalized exactly); and
it can describe the physics at arbitrary low energies and temperatures
(due to the logarithmic discretization). This is important in Wilson's
calculation for the Kondo problem which indeed showed what had been
anticipated by Anderson: the development of a ground state with a 
completely screened impurity (the Fermi-liquid or strong-coupling
fixed point). The crossover to this fixed point occurs at the 
Kondo scale
\begin{equation}
   k_{\rm B} T_{\rm K} = D
       \left( \frac{\Delta}{2U} \right)^{1/2}
 \exp \left( {-\frac{\pi U}{8 \Delta}} \right) \ .
\end{equation}
(This form is valid in the particle-hole symmetric case $\varepsilon_f = - U/2$;
$\Delta$ is defined as
$\Delta = \frac{1}{2}\pi V^2 N(E_{\rm F})$ with $N(E_{\rm F})$ the
density of states of the conduction electrons at the Fermi level).
A sufficiently large ratio $U/\Delta$ can therefore generate
arbitrarily low energy scales.

On the other hand, the NRG method has one main drawback: it is only
applicable to impurity type models and therefore lacks the flexibility
of e.g.\ the Quantum-Monte-Carlo method. A typical example where the
NRG fails is the one-dimensional Hubbard model. This model is
very similar to the semi-infinite chain model of eq. (\ref{eq:chain}), 
but with
constant hopping matrix elements between neighbouring sites and
a Coulomb-repulsion $U$ on each site. One might therefore expect
a similar iterative diagonalization scheme as for the hamiltonian
(\ref{eq:chain}) 
to work for the Hubbard model as well. However, the truncation
scheme (keeping only the lowest lying states)
does not work for a model where the same energy
scales ($U$ and the bandwidth) are added 
at each step of the RG procedure. The low energy spectrum of the
cluster with one additional site now depends on states from
the whole spectrum of energies of the previous iteration.
(A solution to this problem, i.e. finding a truncation scheme
which gives an accurate description of the larger cluster,
is the Density matrix renormalization group method \cite{DMRG}).

There are, fortunately, a lot of interesting impurity models where
the NRG can be applied and where it provided insights into
a variety of physical problems. Non-Fermi liquid behaviour
has been studied in the context of the Two-Channel-Kondo-Model
and related models \cite{TCKM}. The structure of the Non-Fermi liquid fixed
point as well as its stability against various perturbations
has been clarified using the NRG method.

Another example is the quantum phase transition in impurity models
coupling to conduction electrons with a vanishing density of
states at the Fermi level: $\rho_c(\omega)\propto|\omega|^r$.
Here the NRG enables a non-perturbative investigation of both the 
strong-coupling and local moment phases as well as the quantum critical
point seperating these two \cite{pseudo}.

Apart from applying the NRG to generalized impurity models, some
important technical developments have been made during the past
10 - 15 years; most notably the calculation of dynamical properties,
both at zero and finite temperatures \cite{Sak89,Cos94}.

\begin{figure}
\centerline{\sf N=14}
\hspace*{2.8cm}\psfig{figure=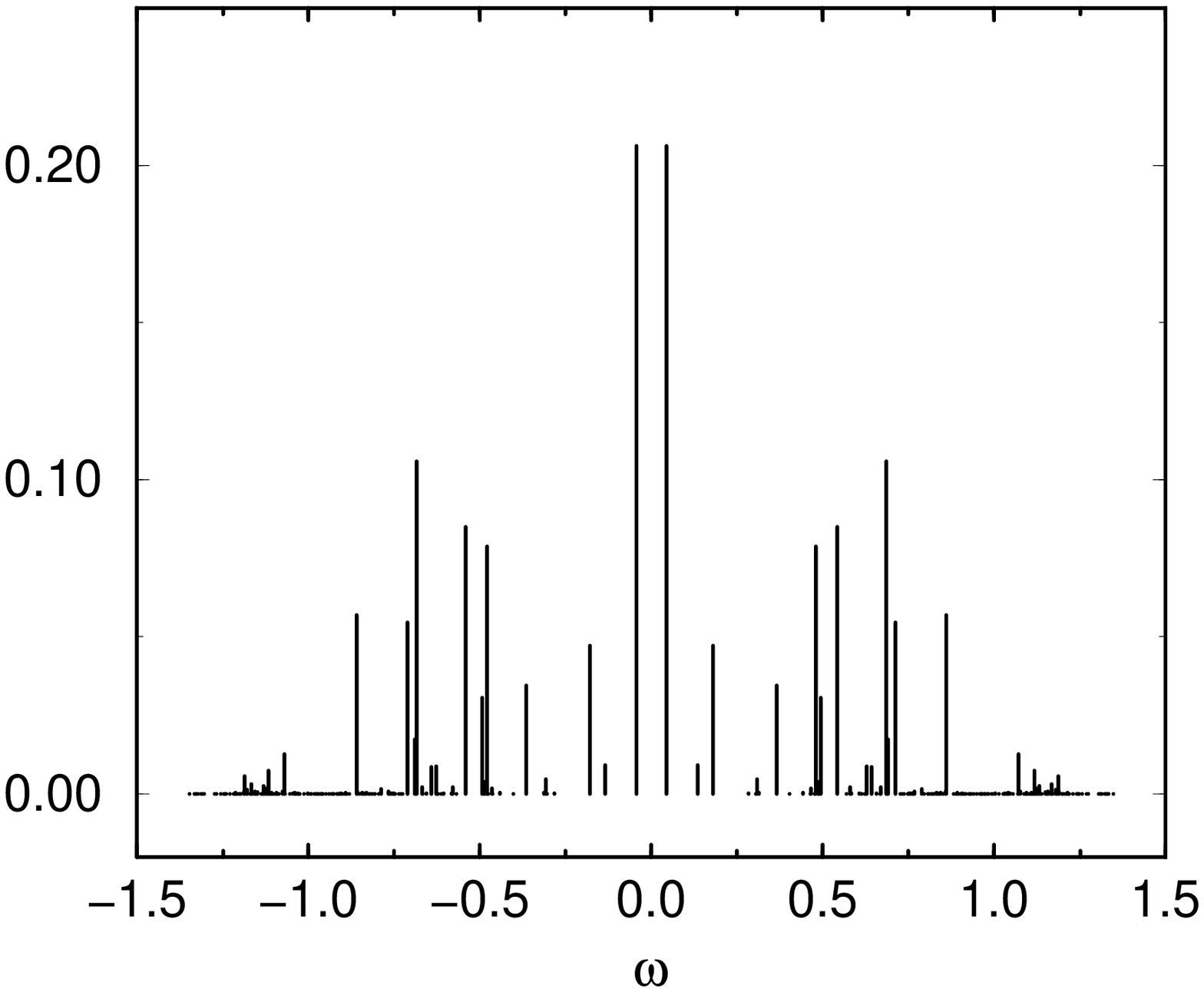,width=6.2cm}
\centerline{\sf N=16}
\hspace*{2.8cm}\psfig{figure=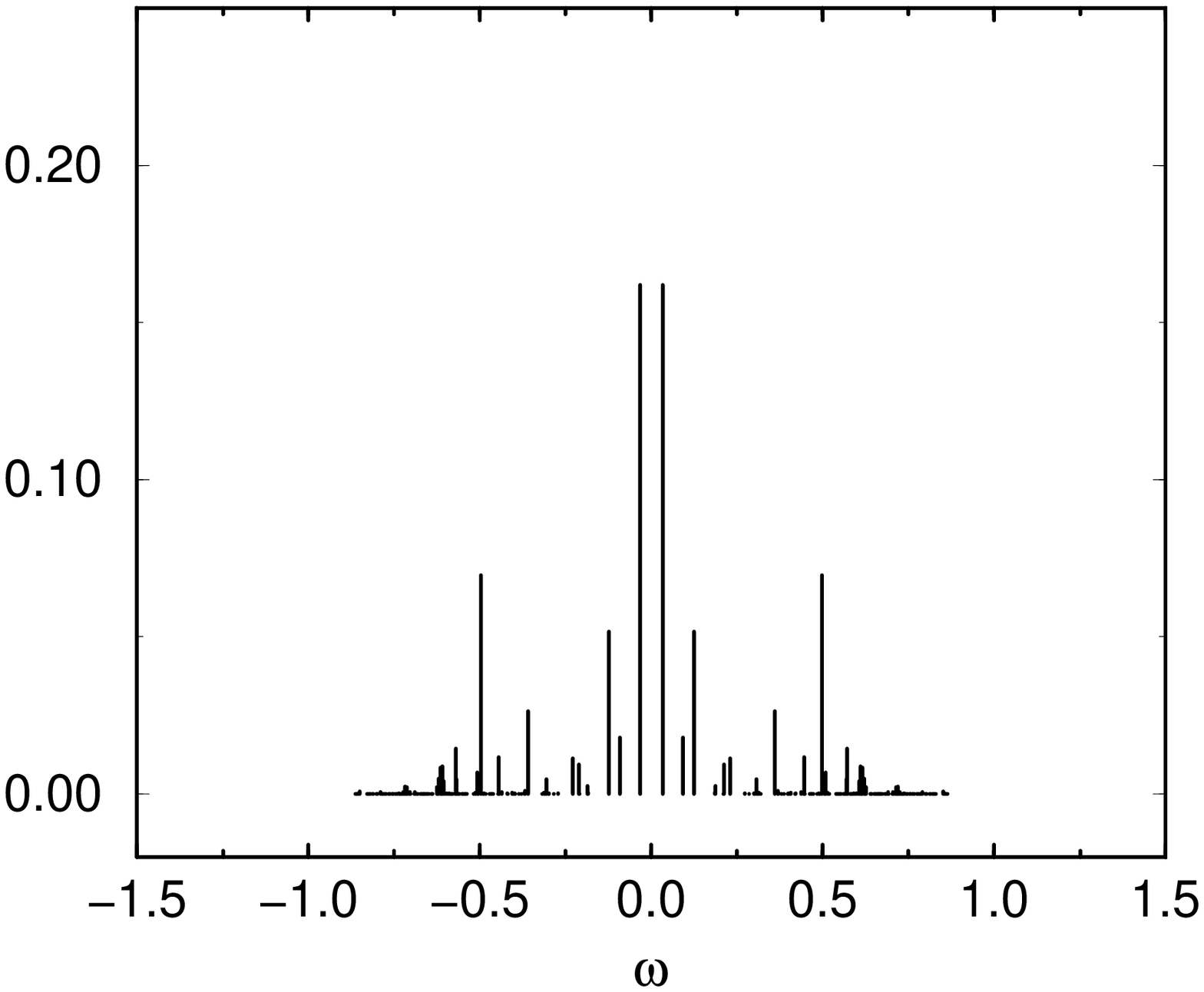,width=6.2cm}
\centerline{\sf N=18}
\hspace*{2.8cm}\psfig{figure=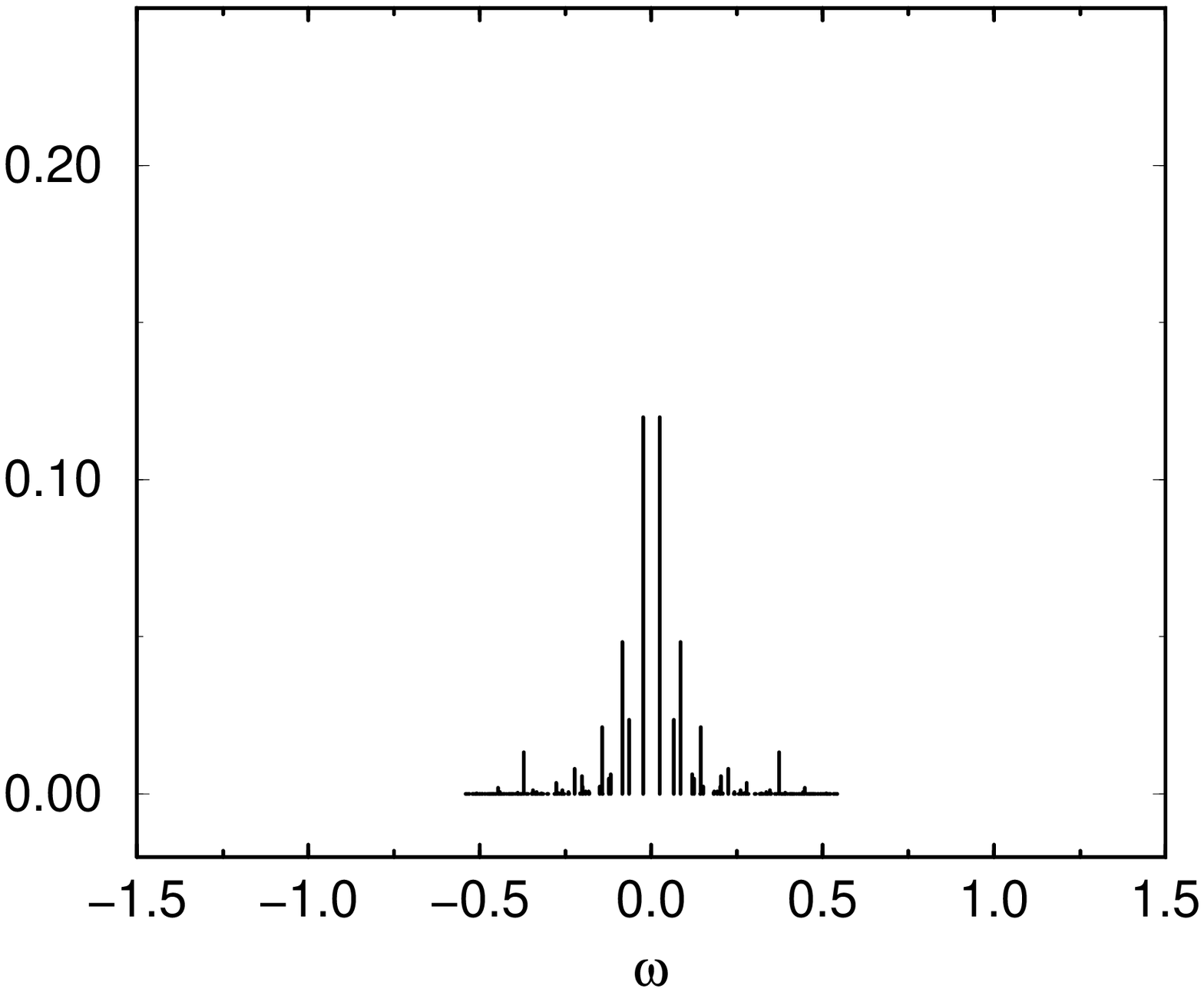,width=6.2cm}
\caption[fig1]{Spectral functions $A_N(\omega)$ for clusters with size $N=14$,
$16$ and $18$. The weight of the $\delta$-peaks is given by
the height of the spikes.}

\label{fig:A_N}

\end{figure}

Let us briefly discuss how to calculate the single-particle
spectral function 
\begin{equation}
  A(\omega) = -\frac{1}{\pi} {\rm Im} G(\omega + i\delta^+)
\ ,\ \ \ {\rm with} \ \ \ \ 
 G(z) = \langle\langle f_\sigma , f_\sigma^\dagger \rangle\rangle_z \ ,
\end{equation}
within the NRG approach. Due to the discreteness of the Hamiltonian, the
spectral function $A(\omega)$ is given by a discrete set of 
$\delta$-peaks and the general expression for finite temperature reads:
\begin{equation}
   A_N(\omega) = \frac{1}{Z_N}\sum_{nm}  
           \bigg\vert   \Big< n \Big\vert f^\dagger_{-1\sigma}
                                         \Big\vert m \Big>
                   \bigg\vert^2
                   \delta \big( \omega -(E_{n} -E_{m}) \big) 
\left( e^{-\beta E_m} + e^{-\beta E_n} \right) \ . 
    \label{eq:Ageneral}
\end{equation}
The index $N$ specifies the iteration number (the cluster size) and for each
$N$ the spectral function is calculated from the matrix elements 
$\big< n \big\vert f^\dagger_{-1\sigma}
                                         \big\vert m \big>$
and the eigenenergies $E_{n},E_{m}$. $Z_N$ is the grand canonical partition function.
Eq. (\ref{eq:Ageneral}) defines the spectral function for each cluster and a
typical result is shown in Fig. \ref{fig:A_N}.

Here, the weight of the $\delta$-peaks in eq. (\ref{eq:Ageneral}) is represented
by the height of the spikes. 
One can clearly see the typical three peak structure from the result of
the 14-site cluster: charge fluctuation peaks centered at 
$\omega\approx \pm 0.7$ ($\omega\approx \pm U/2$) 
and a quasiparticle peak at the Fermi level (here $\omega=0$). However, the
resolution of the quasiparticle peak appears to be rather unsatisfactory:
there is no information on the spectral density below $|\omega| \approx 0.04$.
The advantage of the NRG approach (as compared to e.g. the Exact Diagonalization
technique) is that by successively increasing the length of the chain, one
can extract the information on the spectral density down to arbitrarily low energy scales.
This is seen in the results for the $N=16$ and $N=18$ clusters in Fig. \ref{fig:A_N}.
The necessary truncation of states, as decribed in the previous section, is also
obvious from Fig. \ref{fig:A_N}. There are no excitations for $|\omega| > 0.85$
($|\omega| > 0.45$) in the $N=16$ ($N=18$) cluster, so that the information
on the charge fluctuation peaks is lost for the  $N=16$ and larger clusters. In order
to obtain the spectral density for {\em all} energy scales, the data from
all cluster sizes have to be put together. This means that each cluster size
only provides the information on its relevant energy scale. 

The resulting spectrum
will still be discrete, of course, with the $\delta$-peaks getting closer and
closer together for $\omega\to 0$. It is convenient (both for using the
results in further calculations and for visualizing the distribution of spectral
weight) to broaden the $\delta$-peaks in eq. (\ref{eq:Ageneral}) via
\begin{equation}
  \delta(\omega - \omega_n) \rightarrow 
          \frac{e^{-b^2/4}}{b \omega_n\sqrt{\pi}} \exp \left[ 
      -\frac{(\ln\omega - \ln \omega_n)^2}{b^2} \right] 
\end{equation}
The broadening function is a gaussian on a logarithmic scale with width
$b$. In this way, the broadening takes into account the logarithmic distribution
of the $\delta$-peaks.

\begin{figure}

\hspace*{2.3cm}\psfig{figure=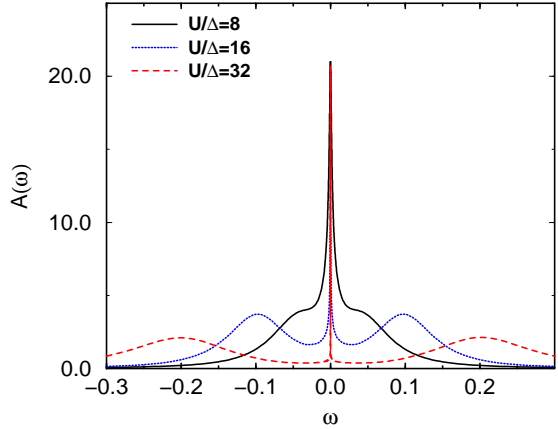,width=8cm}
\caption[fig1]{Spectral functions for the single impurity Anderson
model for various values
  of $U/\Delta$. 
  }

\label{fig:A_SIAM}

\end{figure}

Typical results for the spectral function of the single impurity Anderson
model are shown in Fig. \ref{fig:A_SIAM}. The spectra clearly show the narrowing of
the quasiparticle resonance on increasing the ratio $U/\Delta$ -- corresponding
to the exponential dependence of the low energy scale $T_{\rm K}$ on $U/\Delta$.

Let us now discuss another, very important 
development which made it possible to apply
the NRG method also to lattice models of correlated electrons: the Dynamical
Mean Field Theory (DMFT).

Metzner and Vollhardt \cite{MVPRLdinfty}
 showed that one can define a  non-trivial limit of infinite
spatial dimensions  for lattice fermion models
(such as the Hubbard model). In this limit, the self energy becomes
purely local which allows the mapping of the lattice model onto an effective
single impurity Anderson model. This impurity model has the same structure as in
eq. (\ref{eq:siam}), but the density of states of the conduction band in the impurity
Anderson model has to be determined self-consistently and therefore
acquires some frequency dependence. The NRG can nevertheless
be applied to this case (for details see \cite{BHP}). The first attempts to study
the Hubbard model is the work of Sakai and Kuramoto \cite{Sakai}. The results obtained
later by Bulla, Hewson and Pruschke \cite{BHP} and Bulla \cite{Bulla} will be discussed
in the following section.

\section{NRG results for the Mott-Hubbard metal-insulator transition}

The Mott-Hubbard metal-insulator transition \cite{Mott,BUCH}
is one of the most fascinating phenomena of strongly correlated
electron systems. This transition from a paramagnetic metal to
a paramagnetic insulator is found in various transition metal
oxides, such as $\rm V_2O_3$ doped with Cr \cite{McW}.
The mechanism driving the Mott-Hubbard transition is believed to be the
local Coulomb repulsion $U$ between electrons on a same lattice site, although
the details of the transition should also be influenced by lattice degrees
of freedom. 
Therefore,
the simplest model to investigate the correlation driven metal-insulator
transition is the Hubbard model \cite{Hubbard,Gut,Kan}
\begin{equation}
   H = -t\sum_{<ij>\sigma} (c^\dagger_{i\sigma} c_{j\sigma} +
                   c^\dagger_{j\sigma} c_{i\sigma}) +
         U\sum_i c^\dagger_{i\uparrow} c_{i\uparrow}
            c^\dagger_{i\downarrow} c_{i\downarrow}  ,
\label{eq:H}
\end{equation}
where $c^\dagger_{i\sigma}$ ($c_{i\sigma}$) denote creation
(annihilation) operators for a fermion on site $i$, $t$ is the
hopping matrix element and the sum $\sum_{<ij>}$ is restricted
to nearest neighbors.
Despite its simple structure, the solution of this model turns out to
be an extremely difficult many-body problem. The situation is particularly
complicated near the metal-insulator transition where $U$ and the
bandwidth are
roughly of the same order and perturbative schemes (in $U$ or $t$)
are not applicable.

The DMFT has already been briefly discribed in section 2; this method
enabled a very detailed analysis of the phase diagram of the infinite-dimensional
Hubbard model \cite{Georges,PJF}. 
The nature of the Mott-transition, however, has been the subject of
a lively debate over the past five years (see 
\cite{David,Kehrein,NG,Sch99,Roz99}). This debate focusses
on the existence (or non-existence) of a hysteresis region at very low temperatures.
In such a region, two stable solutions of the DMFT equations should exist:
a metallic and an insulating one. This scenario has been proposed by Georges et
al. based on calculations using the Iterated Perturbation Theory (IPT), Quantum
Monte Carlo and Exact Diagonalization \cite{Georges}.
The validity of this result has been questioned by various authors 
\cite{Kehrein,NG,Sch99}.

\begin{figure}

\hspace*{2.8cm}\psfig{figure=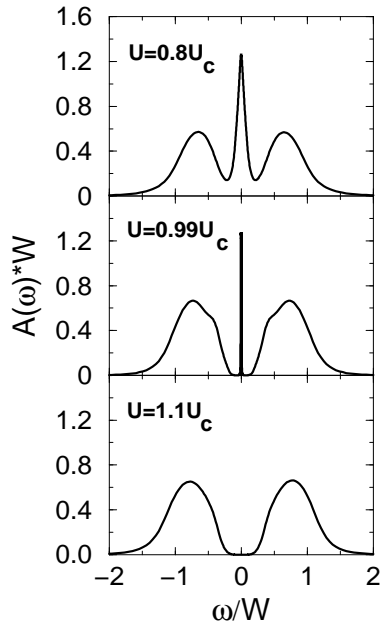,width=11cm}
\caption[fig1]{Spectral functions for Bethe  lattice for various values
  of $U$. A narrow quasiparticle peak develops at the
  Fermi level which vanishes at the critical $U_{\rm c}\approx 1.47W$.
  }

\label{fig:A_T=0}

\end{figure}

Let us now discuss the NRG results for the infinite dimenional Hubbard
model, first of all for $T=0$. The spectral function $A(\omega)$ for the
Bethe lattice is shown in Fig. 
\ref{fig:A_T=0} for  $U\!=\!0.8U_{\rm c}$, $U\!=\!0.99U_{\rm c}$
and $U\!=\!1.1U_{\rm c}$ ($U_{\rm c}\approx 1.47W$, $W$: bandwidth)
 In the 
metallic phase (for large enough values
of $U$) the spectral function shows the typical three-peak structure
with upper and lower Hubbard bands centered at $\pm U/2$
and a quasiparticle
peak at the Fermi level. For $U\!=\!0.99U_{\rm c}$, the quasiparticle peak
in both Bethe and hypercubic 
lattice seems to be isolated (within the numerical accuracy)
from the upper and lower Hubbard
bands, similar to what has been observed in the IPT calculations
for the Bethe lattice
\cite{Georges}. Consequently, the gap appears to open discontinuously
at the critical $U$ (whether the spectral weight between the Hubbard
bands and the quasiparticle peak is exactly zero or very small but finite
cannot be decided with the numerical approach used here).

The quasiparticle peak vanishes at 
$U_{\rm c}\approx  1.47W$
in excellent agreement with the results from the
Projective Self-consistent Method (PSCM) \cite{Georges,Moeller} 
$U_{\rm c}\approx  1.46W$. Coexistence of metallic and insulating solutions
in an interval $U_{\rm c,1}<U<U_{\rm c,2}$ is also found within the NRG approach.
Starting from $U\!=\!0$, the metal
to insulator transition occurs at the critical $U_{\rm c,2}$
with the vanishing of the quasiparticle peak. Starting 
from the insulating side, the insulator
to metal transition happens at $U_{\rm c,1}<U_{\rm c,2}$
(the NRG and IPT give 
$U_{\rm c,1}\approx 1.25W$ for the Bethe lattice).

\begin{figure}

\hspace*{2.0cm}\psfig{figure=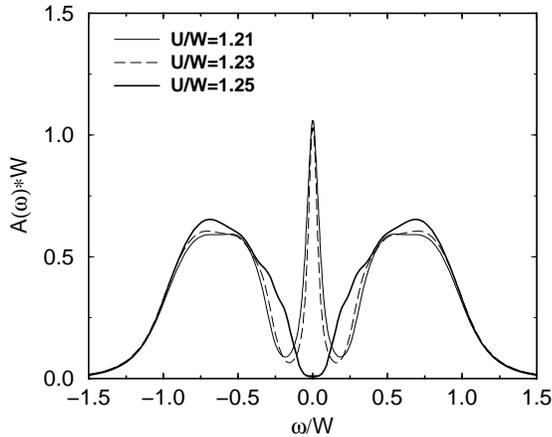,width=8cm}
\caption[fig1]{Spectral functions for $T =0.00625W$ and {\em increasing} 
values of $U$.}

\label{fig:A>0}

\end{figure}

The NRG method for the Hubbard model has only recently been generalized 
at finite temperatures \cite{BCV}. Preliminary results for the spectral 
function are shown in
Fig. \ref{fig:A>0} 
for $T =0.00625W$ and {\em increasing} values of $U$. 
The upper critical
$U$ is given by $U_{\rm c,2} \approx 1.24W$ and the transition at $U_{\rm c,2} $
is of first order, i.e. associated with a transfer of spectral weight.
The `insulator' for $U>U_{\rm c,2}$ does not
develop a full gap (this is only possible for $U\to \infty$ or $T\to 0$), but
the corresponding transport properties in this temperature range will be
certainly insulating-like. 

For this temperature, the NRG again finds two stable solutions
in an interval  $U_{\rm c,1}(T)<U<U_{\rm c,2}(T)$: a metallic one, with
a quasiparticle peak at the Fermi level and an `insulating' one,
with very small spectral weight at the Fermi level (not shown here).  
The exact shape of the hysteresis region has still to be determined
and will be discussed elsewhere \cite{BCV}.

What we have seen in this section is that the NRG-method (together with
the DMFT) can be applied to the infinite-dimensional Hubbard model
and allows a non-perturbative calculation of dynamical properties. The calculations
can be performed for arbitrary interaction strength and temperature, so that
the phase diagram can be (in principle) determined in the full parameter space.

\section{Further developments of the NRG method}

As we have discussed in the previous sections, Wilson's NRG can be applied to two
different classes of problems: impurity models and lattice models 
(the latter ones, however, only within the DMFT).

Concerning impurity models, the NRG has provided important theoretical
insight for a variety of problems and certainly will do so in the future.
In the light of the increasing possibilities of experimental fabrication, new
classes of impurity models are becoming of interest. The behaviour of electrons
in quantum dots, for example, can be interpreted as that of an impurity in
a conduction band (for an application of the NRG method to this problem,
see \cite{quantum_dots}). Magnetic impurities can also serve
as sensors, put into certain materials in a controlled way. Here one might
think of impurities in a correlated host \cite{HBV}, or impurities in a superconducting
or magnetic medium. A lot of theoretical work in applying the NRG method to these
problems still needs to be done.

The second  class of models are lattice models within the DMFT. 
Here, the NRG allows (at least in principle) the calculation
of a large set of experimentally relevant quantities for a wide range of
parameters (especially low temperatures and strong correlation) for a large
class of models. Apart from the application to the Hubbard model which has
been briefly discussed in section 3, the NRG has already been applied to 
the periodic
Anderson model \cite{pam} and to the problem of charge ordering in the extended
Hubbard model \cite{co}. Future work will focus on generalizing the NRG method
to magnetically ordered states and to systems 
with a coupling to (dynamical) phonons.

Of particular interest is the generalization of the NRG to multi-band
models. In this way, the NRG could further extend the range of
applicability of the LDA+DMFT approach \cite{LDAplusDMFT}.
Here, the non-interacting electronic band structure as calculated
by the local density approximation is taken as a starting point,
with the missing correlations introduces via the DMFT.
On a more fundamental level, the basic physics of multi-band models
at low temperatures still needs to be clarified, and again, the NRG
is the obvious choice for investigating such models in the low $T$
and intermediate to large $U$ regime.

The author would like to thank T.\ Costi, D.E.\ Logan, A.C.\ Hewson, W. Hofstetter,
M.\ Potthoff, Th.\ Pruschke, and D. Vollhardt for stimulating discussions
and collaboration over the past few years. Part of this work
was supported by the Deutsche Forschungsgemeinschaft, grant No.\
Bu965-1/1 and by the Sonderforschungsbereich 484.

\end{document}